# Influence of Ga-Ge Doping on the Structural and Electrical Properties of $Li_7La_3Zr_2O_{12}$ Solid Electrolytes For Li ion Battery

**Muktai Aote[a], A. V. Deshpande[a,*], Anuj Khapekar[a], Kajal Parchake[a]**

[a] *Department of Physics, Visvesvaraya National Institute of Technology, South Ambazari Road, Nagpur, 440010, Maharashtra, India*

*****Corresponding Author:**

Dr.(Mrs.)A. V. Deshpande

Department of Physics,

Visvesvaraya National Institute of Technology,

South Ambazari Road, Nagpur

Maharashtra, 440010 (India)

E-mail Address: avdeshpande@phy.vnit.ac.in

Ph.No.: +91-712-280-1251

## Abstract

In order to replace the conventional liquid electrolytes by solid electrolytes, high room temperature ionic conductivity is required. To achieve such high ionic conductivity, the series $Li_{7-3x}Ga_xLa_3Zr_{1.9}Ge_{0.1}O_{12}$ was prepared by solid-state reaction method. The content of Ge was kept constant at 0.10 a.p.f.u, and the Ga has been varied from 0 to 0.40 a.p.f.u. The conducting cubic phase was identified using X-ray diffraction study, whereas the physical and structural studies were perfomed using density measurements and scanning electron microscopy (SEM) analysis, respectively. Electrical conductivity results reveal that the 0.20 Ga ceramic sample possessed the highest room temperature Li ion conductivity of 5.09 x $10^{-4}$ S/cm and minimum activation energy of 0.25 eV. Predominant ionic conduction in 0.20 Ga ceramic sample was confirmed by the DC polarization method. The high room temperature ionic conductivity makes the 0.20 Ga ceramic sample a suitable candidate as a solid electrolyte for all-solid-state lithium-ion batteries (ASSLIBs).

## 1. Introduction

Liquid electrolytes have major issues related to their flammability, dendrite growth formation, and leakage, leading to increased safety concerns for consumers nowadays [1]. Thus, conventional liquid electrolytes are being replaced with solid electrolytes. To be used in all-solid-state batteries (ASSBs), solid electrolytes must have energy and power density comparable to liquid electrolytes. Out of all solid electrolytes, oxide type garnet structured $Li_7La_3Zr_2O_{12}$ (LLZO) proved to have the high electrochemical potential window with stability against Li metal [2] for all solid-state lithium ion batteries (ASSLIB's). Despite good stability with Li metal anode, LLZO fails to achieve the required conductivity range at room temperature.

Garnet-structured LLZO possesses an essential channel for Li-ion migration [3], as the distribution of Li ions in the lattice of LLZO plays an important role in Li-ion conductivity. The initial framework of garnet-structured LLZO comprises $ZrO_6$ octahedra (16 a), a six-fold coordinated structure, and $LaO_8$ dodecahedra (24c), an eight-fold coordinated structure. In this structure, Li ions are distributed over the interstitial sites.[4–7] Within the LLZO lattice, Li-ion occupies different lattice sites: 24d tetrahedral (Li1) site, 48g octahedral (Li2) site, and 96h disordered octahedral (Li3) site [7–9]. Even though disordered octahedral results due to the columbic repulsion between the Li1 and Li2 sites, these tetrahedral and octahedral sites also form the conducting pathways for Li-ion migration within the lattice [5,10].

Initially, LLZO is present in the tetragonal phase ($I4_1/acd$). However, the tetragonal phase of LLZO exhibits low ionic conductivity at room temperature. It has a thermodynamically ordered structure due to fully occupied Li sites, which restricts Li-ion migration within the Li sublattice. On the other hand, the cubic phase of LLZO (Ia-3d), has partially occupied Li sites, resulting in a disordered structure and giving higher ionic conductivity by two orders of magnitude than the tetragonal phase [10,11]. However, forming the cubic phase requires a higher temperature, leading to an unwanted phase that hinders the Li-ion migration pathways and eventually reduces the ionic conductivity [12]. Also, the structural disorder makes it challenging to stabilize the cubic phase at room temperature [1]. The cubic phase can be stabilized by enhancing the Li-ion vacancies within the crystal structure. The Li-ion vacancies can be created by doping supevalent cations in the LLZO lattice, which eventually improves the ionic conductivity by increasing the entropy and reducing the free energy [12]. Additionally, these supervalent cations can also be used as sintering additives, which helps in lowering the sintering temperature required for the formation of cubic phase [1,2].

In the LLZO lattice, supervalent cations can be inserted at any of the sites, including Li, La, and Zr. However, each of the doping sites helps differently in increasing ionic conductivity. The insertion of cations at the La and Zr sites can regulate the possible size of the channel for the migration of Li ions. It also increases the total Li content within the structure of LLZO if the substitution is nonequivalent. On the other hand, the Li site substituted LLZO shows increased Li ion conductivity as it affects the regular distribution of Li ions at various Li sites in the lattice [13].

Initially, the enhancement in ionic conductivity of LLZO was achieved by single doping at any of the Li, La, or Zr sites with the cations like Ge, Ga, Ta, Al, Ca, Bi, Cr, Zn.[11,14–20] The obtained result was attributed to the creation of Li-ion migration pathways, which helped to stabilize the cubic phase and eventually enhanced the Li-ion conductivity.[1] After the satisfying results of a single doping strategy, researchers explored the various co-doped LLZO by simultaneous doping at any of the Li, La, and Zr sites. This study includes the synthesis of Ta – Ge, Al – Ga, Rb –Ta, Ca –Ta, Sr – Ta, Ga –Y doped LLZO [1,10,21–24]. The results of this investigation showed significant impact on the stabilization of conducting cubic phase and improved electrical and structural properties due to the synergistic effect created in garnet LLZO [1,11]. Some of the research findings also report the successful insertion at all the sites, i.e., Li, La, and Zr in LLZO simultaneously. Aote and Deshpande[2] achieved high room temperature Li ion conductivity by introducing Ge, Ca, and Ta at Li, La, and Zr sites respectively. Here, Ca insertion increased the total Li content in LLZO while Ge was used as a sintering aid, and Ta helped in the stabilization of the cubic phase. Even though the results regarding the tridoping method are interesting, very few studies have reported implementing such a strategy [25,26].

All the previous studies with various supervalent cations revealed that Ga effectively improved the Li ion conductivity of LLZO at room temperature. The insertion of Ga at the Li site resulted in the distortion of the LLZO lattice. Li ions occupy the 12a, 12b, 48e Wyckoff sites in this disordered structure. These sites give three different migration routes, resulting in increased ionic conductivity [27,28]. Moreover, the difference in ionic radius of Ga (0.47 Å) and Li (0.59 Å) causes lattice distortion in which it is believed that Ga occupies an octahedral 96h site which is favorable for Li-ion migration [29]. As Ga has proved to help in increasing Li ion conductivity, Ge has proven its ability as a sintering additive [2]. Supervalent Ge as a sintering aid can enhance the sinterability of the ceramic by minimizing the Li loss during cubic phase formation. Many studies have shown that the optimum content of 0.10 atoms per formula unit (a. p. f. u.) of Ge can form the conductive cubic phase at a relatively lower temperature, eventually enhancing the room temperature ionic conductivity [2,29].

Hence, in the present work, the series $Li_{7-3x}Ga_xLa_3Zr_{1.9}Ge_{0.1}O_{12}$ was synthesized using the conventional solid-state reaction method. The 0.10 a.p.f.u. content of Ge was kept constant from our previously reported studies [1,2,11]. It is reported that the fixed content of Ge helps to optimize the total Li-ion content, i.e. 6.4 – 6.6 a.p.f.u. which is required to increase the Li ion conductivity. By keeping Ge constant, the content of Ga has been varied between 0 to 0.40 a.p.f.u. in garnet LLZO and the corresponding synergistic effect on the structure and its electrical properties have been investigated.

## 2. Experimental Work

### 2.1. Material Synthesis

The series $Li_{7-3x}Ga_xLa_3Zr_{1.9}Ge_{0.1}O_{12}$ was prepared by conventional solid-state reaction method with x ranging from 0 to 0.40. The chemicals, namely, $Li_2CO_3$ (Merck, >99.9%), $Ga_2O_3$ (Sigma Aldrich, >99.0%), $La_2O_3$, $ZrO_2$, and $GeO_2$ (Sigma Aldrich, >99.99%) were weighed stoichiometrically and mixed in agate mortar. During the initial mixing process, 10% of excess $Li_2CO_3$ was added to overcome the Li loss occurs during the sintering process. After mixing, the powder was calcined at 900ºC for 8 h using an alumina crucible in a muffle furnace. Once the calcination was done, the powder was again finely crushed. The pellets with a diameter of 10 mm and thickness of around 1.5 mm were made using a hydraulic press under 4 tons of pressure. The pellets were then kept in the mother powder bed and sintered at 1050º C for 8 h. While sintering, the alumina crucible was covered with the alumina lid to avoid air contamination. The prepared samples from the series $Li_{7-3x}Ga_xLa_3Zr_{1.9}Ge_{0.1}O_{12}$ are represented as 0 Ga, 0.10 Ga, 0.20 Ga, 0.30 Ga, and 0.40 Ga for x = 0, 0.10, 0.20, 0.30, 0.40 respectively.

### 2.2. Material Characterizations

Various characterizations have been carried out to investigate the effect of Ge and Ga co-doping on the structural and electrical behavior of garnet LLZO. For the cubic phase identification, the synthesized pellets were grounded into a fine powder and then examined through a RIGAKU X-ray diffractometer. The sample was exposed to Cu-kα radiation which has a wavelength of 1.52 Å. The required information was collected in the range of 10º - 70º by keeping a scan speed of 2º/ min and with a step size of 0.02º. Archimedes' principle was used to calculate the densities of synthesized ceramic samples using the K-15 Classic K-Roy pan balancing instrument. Here, toluene was used as an immersion medium. The information regarding the surface morphology and the presence of elemental composition were investigated through Scanning Electron Microscopy using the JSM-7600F/JEOL instrument. The AC conductivity measurements were carried out using a NOVOCONTROL impedance analyzer within the frequency range of 20 Hz to 20 MHz and temperature range of 25ºC to 150ºC. Furthermore, the DC polarization technique was employed using KEITHLEY 6512 programmable electrometer to ensure the predominance of ionic conduction within the prepared ceramics. The transport number was calculated from the data of DC conductivity. For the AC and DC conductivity measurements, the surface of sintered pellets was coated with silver paste to maintain ohmic contact with the silver electrodes. These electrodes functioned as ion-blocking electrodes in these measurements.

## 3. Results and Discussion

### 3.1. X-ray Diffraction

Fig.1 (a) displays the X-ray diffraction patterns of all the samples of $Li_{7-3x}Ga_xLa_3Zr_{1.9}Ge_{0.1}O_{12}$ series, synthesized using conventional solid-state reaction technique where x is varied from 0 to 0.40. From the X-ray diffraction graph, it is confirmed that the conducting cubic phase (Ia-3d) was achieved for all the Ge and Ga doped LLZO samples. All the obtained X-ray peaks have been indicated with respective (h k l) planes after matching with JCPDS file no. 45.0109 (Represented by red vertical lines). From fig. 1 (a), it can be observed that for 0 Ga and 0.10 Ga small impurity peak was present at around 20º and 27º, corresponding to the $La_2Zr_2O_7$ phase. This Pyrochlore phase is commonly found during the sintering process of garnet LLZO [11,30]. However, as the content of Ga increases from 0.20 to 0.40, the pure cubic phase was observed without any impurity peaks. This confirmed that a minimum of 0.20 Ga is required to stabilize the cubic phase. Also, the peaks related to the 0.20 Ga sample show the highest intensity, sharpness, and crystallinity among all other samples. This obtained result is also supported by the study reported by Xiao Huang et.al.[31]. Xing Xiang et.al.[13] also obtained the cubic phase with the single Ga doping at 1200ºC/18h. However, in the present study, the cubic phase was formed only at 1050ºC/8h. This result can be attributed to the optimum insertion of Ge along with Ga in LLZO, which acts as a sintering aid and helps to stabilize the cubic phase. Moreover, the successful insertion of Ga within the lattice of LLZO can be confirmed from Fig. 1 (b). The peaks in the range from 30º - 35º shifted towards the higher theta values, suggesting the decrease in the lattice parameters. This shifting can be attributed to the difference in ionic radius of Li (0.59 Å) and Ga (0.47 Å).

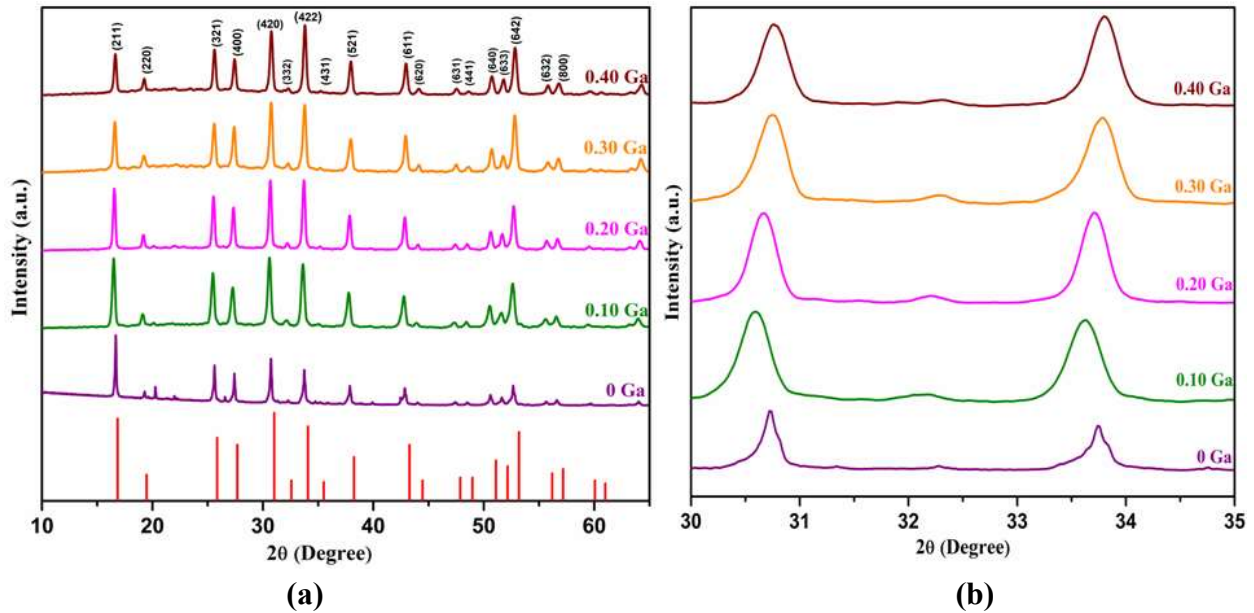

(a)            (b)

**Figure 1: (a) XRD patterns and (b) Shifting of peaks in the series $Li_{7-3x}Ga_xLa_3Zr_{1.9}Ge_{0.1}O_{12}$ with varying x from 0 – 0.40.**

### 3.2. Density Measurement

Archimedes' principle was used to determine the densities of all the synthesized ceramic samples. Toluene was used as an aqueous medium for immersion. The respective values of density and relative density of all the samples of the series $Li_{7-3x}Ga_xLa_3Zr_{1.9}Ge_{0.1}O_{12}$ are mentioned in Table 1. From the table, it can be observed that, with the increase in Ga content, the density increases upto 0.20 Ga. However, further increase in Ga content leads to decrease in density. The highest density was obtained for the 0.20 Ga ceramic sample, with the highest relative density of 94.8%. This enhancement in density can be attributed to the insertion of Ga, as it proved to help in the densification of the LLZO ceramic [10]. It has been reported that the optimum content of Ga helped to improve the sintering ability of LLZO due to the formation of the liquid phase at the sintering temperature [32], which led to the densification of the ceramic sample. However, after the Ga content exceeds its optimum limit of 0.20, there is a slight decrease in density. This may be due to the formation of voids within the sample during the sintering process. The obtained density results can also be correlated with the surface morphological study of the synthesized ceramics, as shown in Fig. 2.

**Table 1: Density and Relative densitiy of $Li_{7-3x}Ga_xLa_3Zr_{1.9}Ge_{0.1}O_{12}$ with varying x from 0 – 0.40.**

| x | Density (g/cm$^3$) | Relative Density (%) |
|---|---|---|
| 0 | 3.93 | 80 |
| 0.10 | 4.58 | 91.6 |
| **0.20** | **4.74** | **94.8** |
| 0.30 | 4.52 | 90.4 |
| 0.40 | 4.43 | 88.6 |

### 3.3. Surface Morphological Study

To get more information about the structural changes that occurred due to the doping of Ga in the series $Li_{7-3x}Ga_xLa_3Zr_{1.9}Ge_{0.1}O_{12}$, the prepared samples were characterized using the SEM technique and the corresponding images are shown in Fig. 2. From the fig. 2 (a), it can be observed that, the surface micrograph of 0 Ga ceramic sample consists of visible voids and open pores. Whereas with the addition of Ga in the sample, the voids were minimized, and the grain size and the compactness increased. The growth in grain size can clearly observed in Fig. 2 (c), corresponding to the 0.20 Ga ceramic sample. All the grains are very well connected with the neighboring grains giving the dense morphology for the synthesized sample. The compact structure of 0.20 Ga sample is in agreement with the highest density result of this sample. This enhancement in grain growth can be attributed to the optimum insertion of Ga along with Ge into the lattice of LLZO. As per the study reported by Shufeng Song et.al.[33], Ga helps to form the transgranular grains, eventually leading to lower grain boundary resistance and higher grain boundary strength. The average particle size distribution for the 0.20 Ga ceramic sample is shown in Fig. 3. The average particle size was found to be 7.4 ± 0.05 μm. However, as the Ga content exceeds the limit, there is formation of pores within the structure, which can be observed from Fig. 2 (d and e). Previous studies suggested that the formation of pores may be due to the evaporation of Li during the sintering process [1,10,34]. This result is also supported by the measured density values for the respective ceramic samples, as listed in Table 1.

Fig. 4 represents the elemental mapping along with EDS spectra of all the constituent elements of the 0.20 Ga ceramic sample. The figure confirms the presence and uniform distribution of all the elements, i.e., La, Zr, Ge, Ga over the surface of the sample. Here, Li was not detected due to the lower energy. It also supports the successful inclusion of Ga and Ge in the lattice of LLZO, as discussed in the X-ray diffraction section and shown in Fig. 1 (b), which depicts the shifting in peaks.

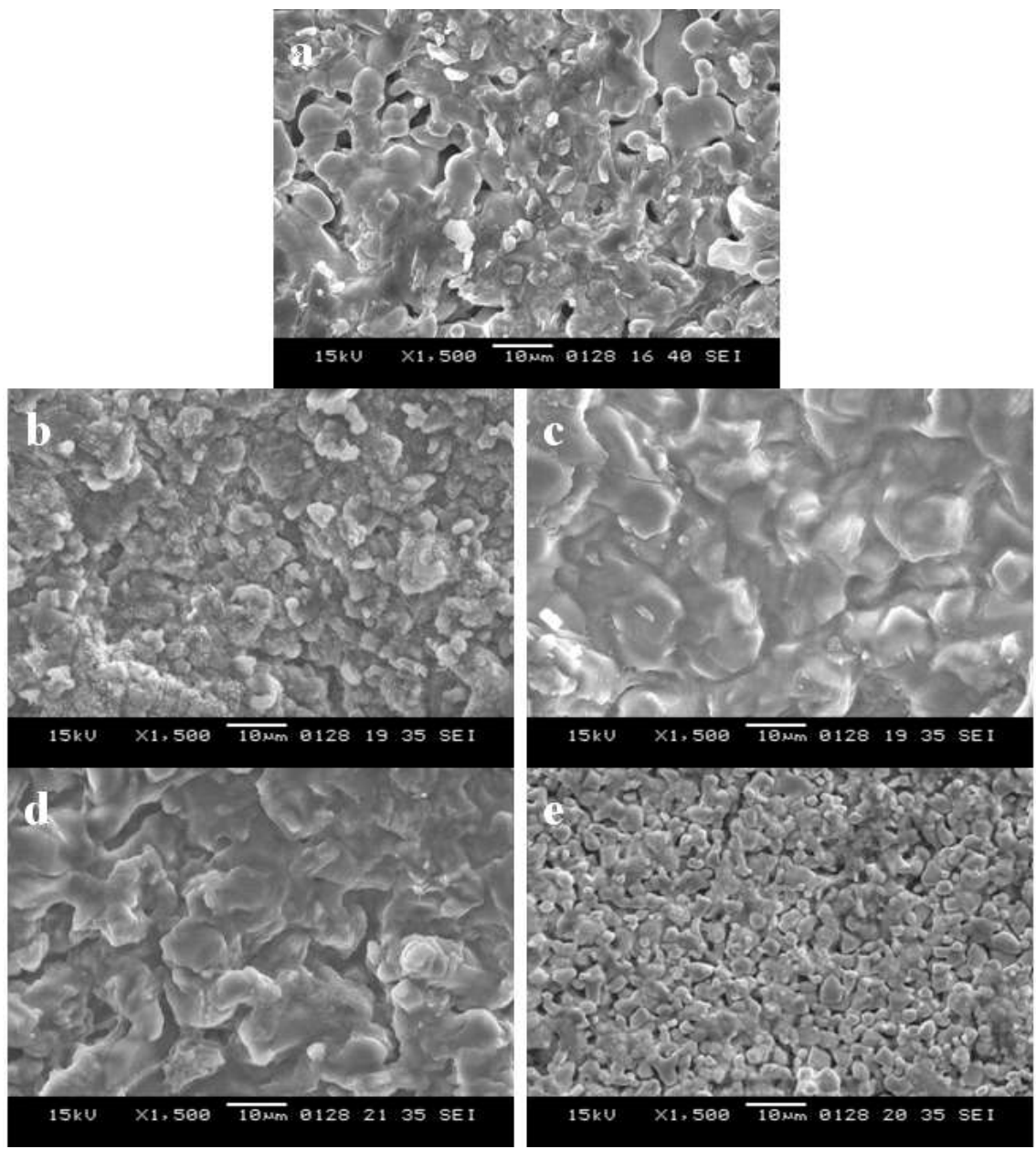

**Figure 2: Surface Morphology images of Li$_{7-3x}$Ga$_x$La$_3$Zr$_{1.9}$Ge$_{0.1}$O$_{12}$ with x = a) 0, b) 0.10, c) 0.20, d) 0.30, e) 0.40.**

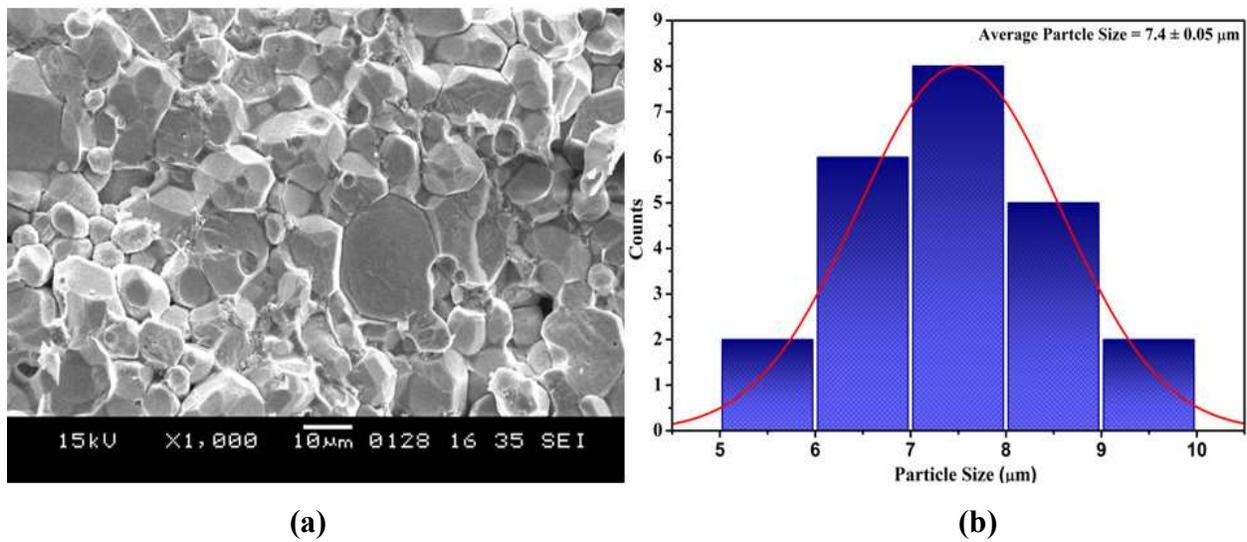

**Figure 3: (a) Cross-sectional Image and (b) Average particle size distribution histogram of 0.20 Ga ceramic.**

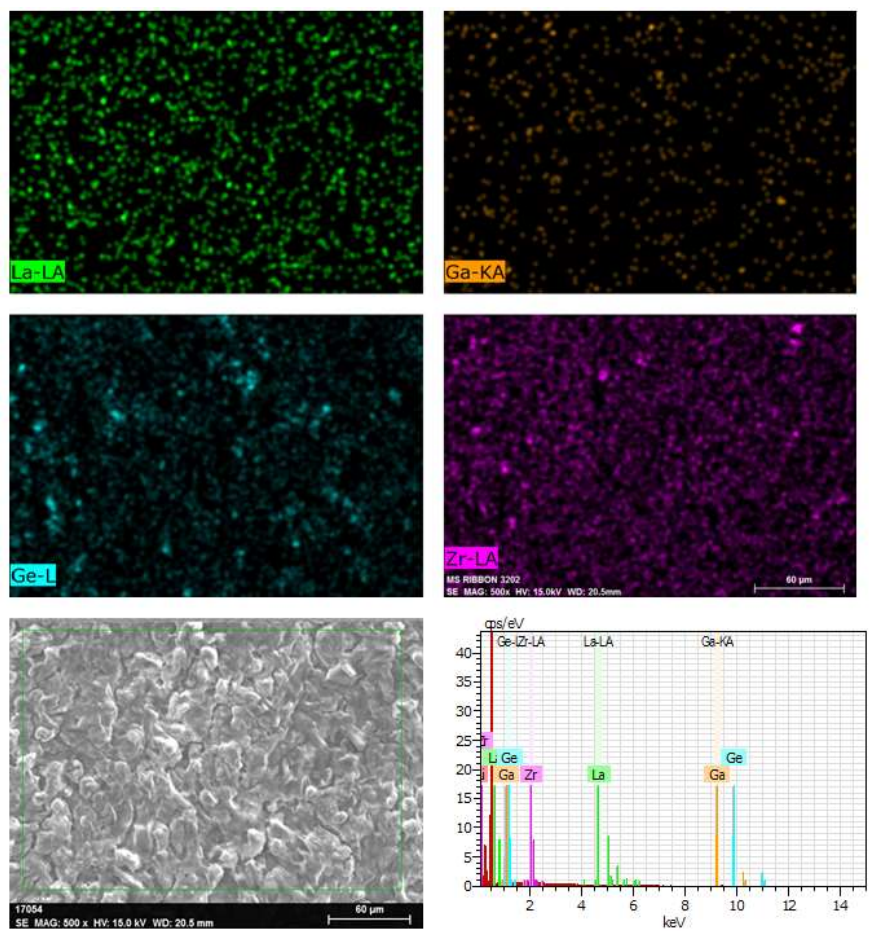

**Figure 4: Elemental mapping and EDS spectra of 0.20 Ga ceramic sample.**

### 3.4. AC conductivity Studies

#### 3.4.1. Impedance Plots

The Nyquist plots at room temperature for all the synthesized ceramic samples of the series $Li_{7-3x}Ga_xLa_3Zr_{1.9}Ge_{0.1}O_{12}$ (x = 0 – 0.40) have been displayed in Fig. 5 (a). The semicircle corresponding to each sample gives the value of the respective impedance of that particular sample. The nature of the graph primarily depicts the ionic conduction within the samples, as it possessed the semicircle intersecting real z-axis in the higher frequency region and the tail in the lower frequency region [35]. The impedance due to the grain and grain boundary of the sample is generally described by the single semicircle obtained in the higher frequency region, whereas the tail in the lower frequency region represents the ion blocking nature of the Ag electrodes [1,2]. The formula $\sigma_{total} = t/RA$ was used to calculate the ionic conductivity of the prepared ceramic samples. Here, $\sigma_{total}$ is the ionic conductivity, $t$ is the thickness and $R$ is the resistance of the sample and $A$ is the area of the electrode. From the inset graph of Fig. 5 (a), it can be observed that the maximum resistance is offered by the 0 Ga sample. As the content of Ga increased, the semicircle shifts towards lower impedance values. The maximum ionic conductivity of 5.09 x $10^{-4}$ S/cm was achieved for the 0.20 Ga ceramic sample which is greater than 0 Ga by two orders of magnitude. Fig. 5 (b) shows the fitted Nyquist plot for 0.20 Ga ceramic sample with the corresponding equivalent circuit. This enhancement in ionic conductivity can be attributed to the optimum content of Ga, which minimizes the length of Li-ion migration pathways and helps the Li ions in hopping through interstitial sites within the garnet LLZO lattice [13,36]. The result of the highest ionic conductivity can also be correlated with the dense microstructure and highest relative density of the 0.20 Ga ceramic sample, which also plays an essential role in the Li-ion conductivity of LLZO. However, with the further increment in Ga content beyond 0.20 a.p.f.u., a slight decrease in conductivity can be observed as the semicircle shifts towards the higher impedance values. This might be due to the presence of voids within the microstructure, as seen in Fig. 2 (d and e) which eventually decreases the relative density and hinders the Li-ion migration pathways [1]. Thus, from the above discussion, it can be confirmed that optimum Ga content of 0.20 a.p.f.u is helpful in the enhancement of Li ion conductivity along with the insertion of Ge (0.10 a.p.f.u.) as it maintained the optimum Li content, required for achieving high Li ion conductivity [11,29]. The obtained result with 0.20 Ga is in well agreement with the previously reported study by Chao Chen et. al.[10].

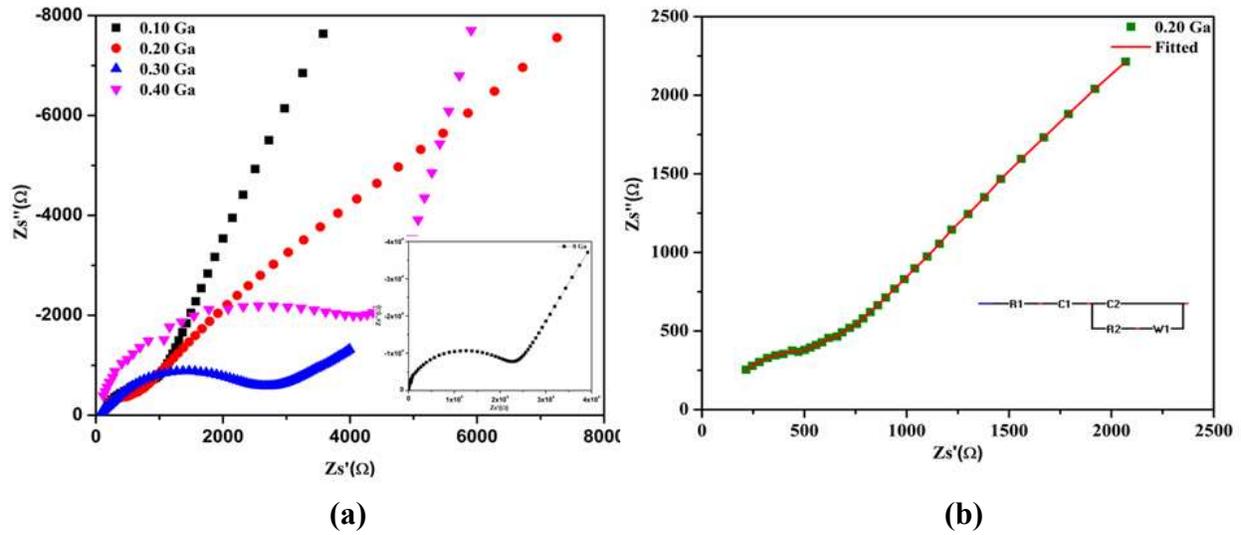

**Figure 5: (a) Nyquist plots of $Li_{7-3x}Ga_xLa_3Zr_{1.9}Ge_{0.1}O_{12}$ with x ranging from 0 – 0.40 at 25ºC. (b) Fitted spectra with an equivalent circuit for 0.20 Ga ceramic sample.**

*3.4.2. Arrhenius Plots*

The temperature dependence behavior of the series $Li_{7-3x}Ga_xLa_3Zr_{1.9}Ge_{0.1}O_{12}$ with x ranging from 0 to 0.40 can be observed from the Arrhenius plots given in Fig. 6 (a). The data were collected in the temperature range from 25ºC to 150ºC. From the fig. 6 (a), it can be observed that all the synthesized samples followed the Arrhenius behavior. The activation energy for these samples was calculated using the Arrhenius equation, $\sigma(T) = \sigma_0 \exp(-E_a/K_B T)$. Here, $\sigma$ is the conductivity, $\sigma_0$ is the pre-exponential factor, $E_a$ is the activation energy, $K_B$ is the Boltzmann constant, and $T$ is the temperature (K). Fig. 6 (a) shows that the minimum activation energy of 0.25 eV was obtained for the 0.20 Ga ceramic sample, which offered the maximum room temperature ionic conductivity of 5.09 x $10^{-4}$ S/cm. The activation energy and ionic conductivity of all the synthesized ceramic samples is given in Table 2. As the Ga content increased from 0 to 0.20 (a.p.f.u.), the activation energy decreased. Whereas, with the further increase in Ga content, it increased. This can also be seen from fig. 6 (b), which represents the variation in activation energy and ionic conductivity with the varying content of Ga. The obtained result can be attributed to the insertion of Ga at the Li site, which created the Li-ion vacancies and disrupted the long-range order of Li site occupancy by contracting the solid phase structure [36]. Moreover, according to previous research, the obtained minimum activation energy can also be due to the insertion of Ge along with Ga as it helps to synthesize the conducting cubic phase at a relatively lower sintering temperature, thus preventing the Li loss from the sample [29].

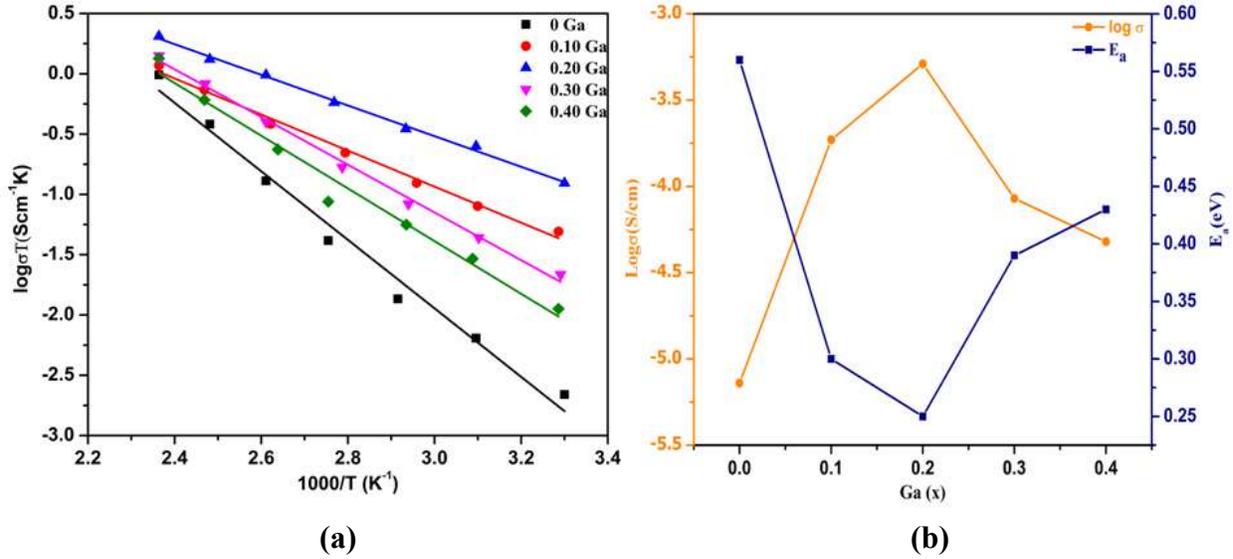

**Figure 6:** (a) Arrhenius plots for $Li_{7-3x}Ga_xLa_3Zr_{1.9}Ge_{0.1}O_{12}$ with x ranging from 0 – 0.40.
(b) Variation in Ionic conductivity and Activation energy with the content of Ga(x).

**Table 2:** Values of Ionic conductivity at 25°C and Activation energy of all the ceramic samples of $Li_{7-3x}Ga_xLa_3Zr_{1.9}Ge_{0.1}O_{12}$ series.

| x | Ionic conductivity (S/cm) | Activation Energy (eV) |
|---|---|---|
| 0 | 8.6 x 10$^{-6}$ | 0.56 |
| 0.10 | 1.85 x 10$^{-4}$ | 0.30 |
| **0.20** | **5.09 x 10$^{-4}$** | **0.25** |
| 0.30 | 8.4 x 10$^{-5}$ | 0.39 |
| 0.40 | 4.8 x 10$^{-5}$ | 0.43 |

### *3.5. DC Conductivity Study*

The electronic conductivity was calculated for all the ceramic sample in series $Li_{7-3x}Ga_xLa_3Zr_{1.9}Ge_{0.1}O_{12}$ using the DC polarization technique. Fig. 7 shows the DC conductivity graph for the 0.20 Ga sample. Silver paste was applied on both faces of the pellet, and the arrangement was done exactly as in the AC conductivity studies. The constant voltage of 1V was applied and the corresponding current through the sample was measured. From the fig.7, it can be observed that, as time passes, the current through the sample decreses and remains almost constant which is due to electrons only [1,11]. However, to ensure the predominant nature of ionic conductivity over electronic conductivity in 0.20 Ga ceramic sample, the ionic transport number was calculated using the formula as $t_i = (\sigma_{total} - \sigma_e)/\sigma_{total}$. It was found to be > 0.999, confirming the negligible electronic conductivity in the 0.20 Ga ceramic sample [2,10].

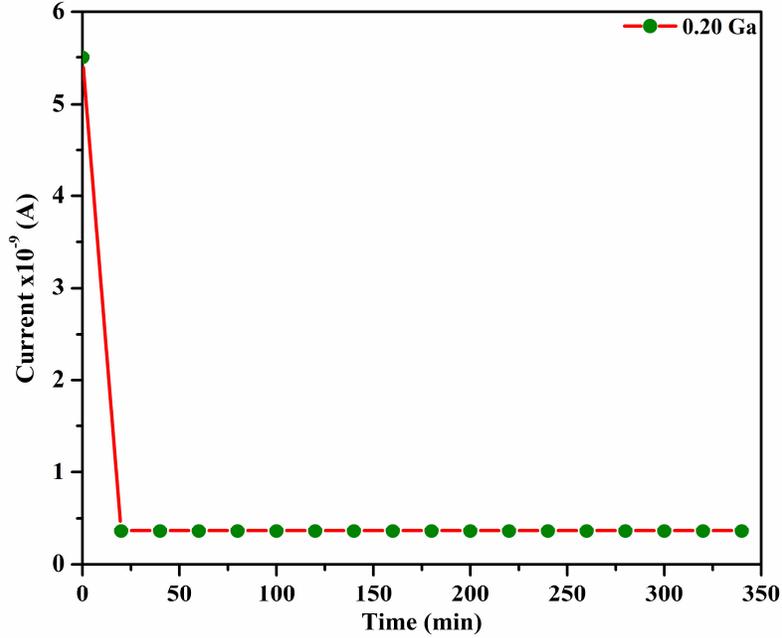

**Figure 7: DC conductivity graph of 0.20 Ga ceramic sample.**

## 4. Conclusions

The conventional solid-state reaction method was used to synthesize ceramic samples in the series $Li_{7-3x}Ga_xLa_3Zr_{1.9}Ge_{0.1}O_{12}$ with Ga(x) content ranging from 0 – 0.40. With the doping of Ge along with Ga as a sintering aid, all synthesized samples possessed the conductive cubic phase at a relatively lower sintering temperature of 1050º C. X-ray diffraction pattern confirmed that the minimum content of Ga is 0.20 a.p.f.u. for the formation of cubic phase without any impurity phase. The 0.20 Ga ceramic sample has the highest relative density of 94.8%, which was also confirmed by the surface morphology images. 0.20 Ga also helped in grain growth, giving the average particle size of 7.4 ± 0.05 μm. The insertion of Ga in the LLZO lattice minimized the Li-ion migration channel, leading to the highest ionic conductivity of 5.09 x $10^{-4}$ S/cm at room temperature. Furthermore, the Li-ion vacancies created due to the Ga helped to minimize the activation energy. The minimum activation energy of 0.25 eV and the highest ionic conductivity was exhibited by 0.20 Ga ceramic sample among all the samples of the series. The negligible electronic conductivity was ensured from the ionic transport number measurement, which was > 0.999 for the 0.20 Ga ceramic sample. All these results for 0.20 Ga ceramic sample makes it a prominent candidate as a solid electrolyte for the future of all-solid-state lithium-ion batteries (ASSLIBs).

**Author Contribution**

**Muktai Aote:** Conceptualization, Methodology, Analysis of the results, Writing original Draft, Editing

**Anuj Khapekar, Kajal Parchake :** Methodology, Formal Analysis, Investigation.

**A.V.Deshpande:** Supervision, review and editing of the draft, Conceptualization

**Funding Sources**

This research did not receive any grant from any funding agencies.

**Acknowledgement**

One of the Author like to acknowledge VNIT Nagpur for providing the Ph.D. Fellowship. Authors also wish to aknowledge XRD facility supported by DST FIST project number SR/FST/PSI/2017/5(C) provided by Department of Physics, VNIT, Nagpur.